# A complex network approach to time series analysis with application in diagnosis of neuromuscular disorders

Samaneh Samiei[1], Nasser Ghadiri[1], Behnaz Ansari[2]

**Abstract**

Electromyography (EMG) refers to a biomedical signal indicating neuromuscular activity and muscle morphology. Experts accurately diagnose neuromuscular disorders using this time series. Modern data analysis techniques have recently led to introducing novel approaches for mapping time series data to graphs and complex networks with applications in diverse fields, including medicine. The resulting networks develop a completely different visual acuity that can be used to complement physician findings of time series. This can lead to a more enriched analysis, reduced error, more accurate diagnosis of the disease, and increased accuracy and speed of the treatment process.

The mapping process may cause the loss of essential data from the time series and not retain all the time series features. As a result, achieving an approach that can provide a good representation of the time series while maintaining essential features is crucial. This paper proposes a new approach to network development named GraphTS to overcome the limited accuracy of existing methods through EMG time series using the visibility graph method. For this purpose, EMG signals are pre-processed and mapped to a complex network by a standard visibility graph algorithm. The resulting networks can differentiate between healthy and patient samples. In the next step, the properties of the developed networks are given in the form of a feature matrix as input to classifiers after extracting optimal features. Performance evaluation of the proposed approach with deep neural network shows 99.30% accuracy for training data and 99.18% for test data. Therefore, in addition to enriched network representation and covering the features of time series for healthy, myopathy, and neuropathy EMG, the proposed technique improves accuracy, precision, recall, and F-score.

[1] *Department of Electrical and Computer Engineering, Isfahan University of Technology, Isfahan 84156-83111, Iran* (e-mails: `s.samiei@ec.iut.ac.ir, nghadiri@iut.ac.ir`)

[2] *Department of Neurology, Isfahan University of Medical Sciences, Isfahan, Iran* (email: `ansaribehnaz@yahoo.com`)

# 1   Introduction

More than 150 types of neuromuscular diseases have been identified to date. This includes disorders of the nerve fibers and muscle tissue of the body [1]. Myopathy and neuropathy are two critical types of neuromuscular disorders, which are referred to as neurological disorders in skeletal muscle and disorders of muscle-controlling nerves, respectively [2-4]. EMG signals have been a primary source of data for diagnosing neuromuscular disorders over the last two decades [4, 5]. This includes signals recorded on electrocardiography, electroencephalography, electromyography. Analyzing these types of signals can lead to an accurate and correct distinction between myopathy and neuropathy [6]. In other words, this is a crucial process of diagnosis and treatment of disorders [7].

EMG signals are recorded using electrodes placed on the surface of the skin or inserted into the skin with a needle [8-10]. Signal recording with noise due to various factors such as incorrect electrode positioning or patient movements during the test can lead to misdiagnosis [1]. Therefore, maximum effort should be made to reduce noise in the accurate analysis of these signals. Advanced data analysis techniques have recently led to developing several methods for mapping time series to complex networks and then applying network criteria to discover hidden patterns of time series. These techniques allow experts to summarize the time series features into regular and comprehensive criteria by reducing the noise in the signal and focusing on the initial values [11].

Constructing a complex network from a time series would provide new visual insight into the signal structure for the specialist. This will potentially complement the findings, reduce medical errors, increase accuracy, and fast-track the treatment process by more accurately diagnosing the disease. A network mapping may not preserve all the features of the signal. Therefore, developing new criteria for converting and representing time series to a complex network while preserving all its features is challenging.

Various time series lead to networks with distinct topological properties, so time series features can be better obtained by analyzing these networks [11, 12]. The approaches used in the studies to achieve this goal include three primary classes: proximity networks [12, 13], visibility graph [14, 15], and transition networks [16].

Existing approaches have high computational complexity. Yet, they provide limited accuracy as required to summarize the time series criteria into a complex network and classify diseases without considering the effects of noise on the classification results. Moreover, the datasets in some works are not explicitly stated, and the methods are opaque with little or no representations of the developed networks.

In this paper, we propose GraphTS as a novel method for building a complex network from the time series. First, we prepare and process the three groups' signals consisting of healthy, myopathy, and neuropathy. The signals are windowed for accurate identification and investigation of the behavior of signals. Then, the linear envelope of the signal is applied to reduce the noise effect and focus on the initial values considering the fluctuating patterns of each group. Finally, a complex network is developed through time series using the standard visibility graph algorithm. Visualization of the networks resulting from the proposed approach is a valuable tool for experts in this field. The network features are also fed into machine learning architectures to help in automating the diagnosis. The main contributions of the proposed method are as follows:

- Obtaining the linear envelope of the signal for building an advanced network with less noise through time series
- Providing the visualization of the networks to enrich the expert's view of the signal
- Network feature extraction and machine learning to investigate the signal based on fundamental differences between healthy, myopathy, and neuropathy groups
- Discovering hidden patterns according to the fluctuating patterns of healthy, myopathy, and neuropathy group

## 2   Basic Concepts

EMG is a biomedical signal received as a set of electrical signals from any organ that shows physical changes [17, 18]. This signal is usually a function of time and can be described in terms of scope, frequency, and phase. An EMG signal is a biomedical signal that shows the electrical currents generated during muscle contraction and stimulation in the form of a time series and represents neuromuscular activity [1, 18].

Neuromuscular Disorder (NMD) refers to diseases that affect any part of a nerve or muscle, including motor neurons, neuromuscular junctions, and muscle tissue [1]. Generally, neuromuscular disorders include a wide range of diseases affecting the peripheral nervous system, including all the motor and sensory nerves that connect the brain and spinal cord to the rest of the body. Neuromuscular disorders are inherited and acquired and are divided into several main types based on the mechanism of disease onset, age at onset, type of heredity, and disease progression [19].

Myopathy is independent of any nervous system disorder and describes a group of diseases that directly affect muscle tissue. Neuropathy refers to a group of diseases that cause disorders in muscle-controlling nerves [2-4]. Neuro means neuron, and pathy means injury. Damage to nerve fibers for a variety of reasons is called neuropathy [19]. This is a neuromuscular disorder, which is any disorder in the path of a motor unit, including motor nerve fibers that start in the brain and spinal cord, where these nerve fibers reach the body's muscles [2, 19].

As an illustrated example, three types of EMG signals received from three samples, including the healthy group, the group with myopathy, and the group with neuropathy, can be seen in Fig. 1 [20].

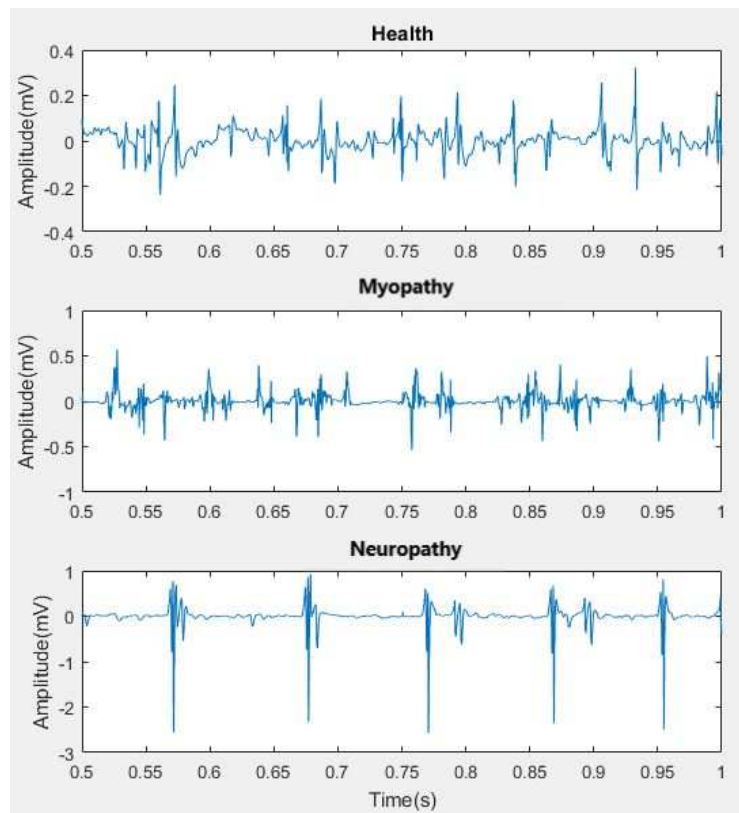

Figure 1. EMG signals

The signal is specified for all three samples on the horizontal axis in an interval of {0.5, 1} seconds, on the vertical axis for the healthy group in an interval of {-0.4, 0.4} microvolts, for the group with myopathy in an interval of {-1, 1} microvolts, and for the group with neuropathy in an interval of {-3, 1} microvolts.

A complex network is an advanced data structure in computer science, often represented as a graph consisting of two primary parts: a finite set of vertices or nodes (V) and edges (*E*). A graph is shown with the symbol *G = (V, E)*. Each vertex of the graph is defined as a numerical index as p = *1, .... N*, and each edge is indicated by an ordered pair of connected elements (*p, q*) [21-23].

## 3 Related work

Diagnosing and classifying diseases can be divided into two primary categories, as described below with some related methods.

### 3.1 Methods based on time series

In this section, we present the literature on diagnosing and classifying diseases in which the time series is *directly* analyzed and diagnosed. In a study [24], the public dataset [25] was used. In the preprocessing section, empirical mode decomposition (EMD) was applied to eliminate the base noise. The feature extraction process was then utilized to detect the feature vector that contains effective information hidden in the signal. Features such as mean, kurtosis, energy, shape factor, and several other features were given to the support vector machine (SVM) classifier and logistic regression after extraction.

By Swaroop et al. [26], the critical role of EMG signal analysis in biomedical engineering and the detection of neuromuscular disorders was emphasized, and this signal was examined. This study's samples were related to three men, including healthy, myopathy, and neuropathy groups. The dataset was from the Department of Neurology, Harvard Medical School. In the proposed approach, fluctuations occurring in high amplitudes were implemented to distinguish between myopathy and neuropathy. Zhang et al. [27] presented a method that, EMG signal processing and analysis were divided into two principal parts: feature extraction and model learning and model-based testing. This study employed datasets provided by the University of Essex in the United Kingdom in the UCI Machine Learning Database. According to what was introduced as the proposed method of the mentioned study, four features in the time domain were extracted from the dataset. The extracted features were then given as input to the deep belief network (DBN).

## 3.2 Methods based on complex network analysis developed through time series (Visibility graph)

To develop a complex network through EEG signals, Wang et al. [7] used three approaches, a standard, and horizontal visibility graph method and a method derived from these two methods. Wang's research team analyzed EEG signals. The data set was related to 29 patients.

In a study [28], while reminding that EMG is a complex and non-stationary signal, 512 time-points of the EMG signal were processed and transformed into a complex network with a standard visibility graph algorithm. The structural features of the developed network, along with the connections and distances between nodes, were then examined. These examinations lead to an indirect binary adjacency matrix. In the next step, values of the network features were given to the support vector machine classifier as a feature vector. In another study [29], time series were first processed with 512 sampled time points. A complex network with interconnected nodes was then obtained using the standard visibility graph algorithm. The developed network had an adjacency matrix through which the interconnected nodes had a visibility graph weight. Therefore, the signal features that were converted into a network were obtained using the visibility graph weight matrix. After extraction, the features were given as a vector as input to the multilayer neural network classifier. Artameeyanant et al. [30], proposed a feature extraction technique based on a standard visibility graph algorithm using weight normalization to diagnose myopathy and ALS. The data sets used in this study were taken from two databases [25] and [20].

The diagnosis of brain disorders using EEG signals is studied in [31]. A new technique was proposed to diagnose seizures in patients with epilepsy through EEG signals using the features extracted from the weighted visibility graph [32]. In this study, the EEG dataset recorded and published by the Epilepsy Center at the University of Bonn, Germany, was analyzed.

Mathur et al. [33] suggested an epilepsy diagnosis technique based on the EEG database that was published by the Epilepsy Center at the University of Bonn, Germany. This database contains 100 samples of each group of EEG signals. Initially, 2,000 time-points from the EEG time series were transformed into a complex network according to a visibility graph algorithm. The weight was then assigned to the edges of this network using Gaussian kernel functions. Finally, the proposed model classified healthy and epileptic classes with 100% accuracy. Besides, in the study [34], an approach was provided in which sleep stages were classified based on the amplitude features extracted from the visibility graphs obtained from the single-channel EEG signal dataset with the SVM learning model.

# 4 Methodology

In our proposed GraphTS method, the EMG dataset is first examined. Noise is removed from the signal by a two-step preprocessing that involves *signal windowing* and finding its *linear envelope*, focusing on the signal's initial values. The time series is then mapped to a *complex network* using a standard *visibility graph* algorithm. In the next step, complex networks are built from the EMG time series data. The network model will contain essential information about the time series that form the network features.

The represented features are then extracted, and a matrix called the features matrix is formed. This matrix is given as input to the classifiers. For some classification models, the feature matrix is statistically verified by analysis of variance before applying it to the classifiers. Fig. 2 represents an overview of the steps of the proposed GraphTS approach, described below.

## 4.1 Data preprocessing

Neuroscientists typically examine signals in the time domain, extract specific windows of the signal, and classify signals by creating specific patterns [30]. EMG signals in healthy, myopathy, and neuropathy groups are quite different from each other and have unique features, which are summarized below:

- **Healthy**: They have at least four phases (the phase in which MUAP intersects the baseline), a length of 5 to 15 seconds, and an amplitude of 100 microvolts to 2 millivolts;
- **Myopathy**: They are polyphasic, low-amplitude, and short in length;
- **Neuropathy**: They are polyphasic, high-amplitude, and long.

Based on the above categories, the windowing operation in the preprocessing of healthy, myopathy, and neuropathy groups is carried out according to the type of signal in each group. Then the operation of finding the linear envelope of each signal is performed.

### 4.1.1 Windowing

An initial step in processing non-stationary signals such as vital and speech signals whose frequency characteristics change over time is windowing [35]. When examining an EMG signal, it is segmented into short windows, and each window is analyzed separately. The windowing process is designed so that a peak of the window is detected, and other parts of the window that have noise will be removed.

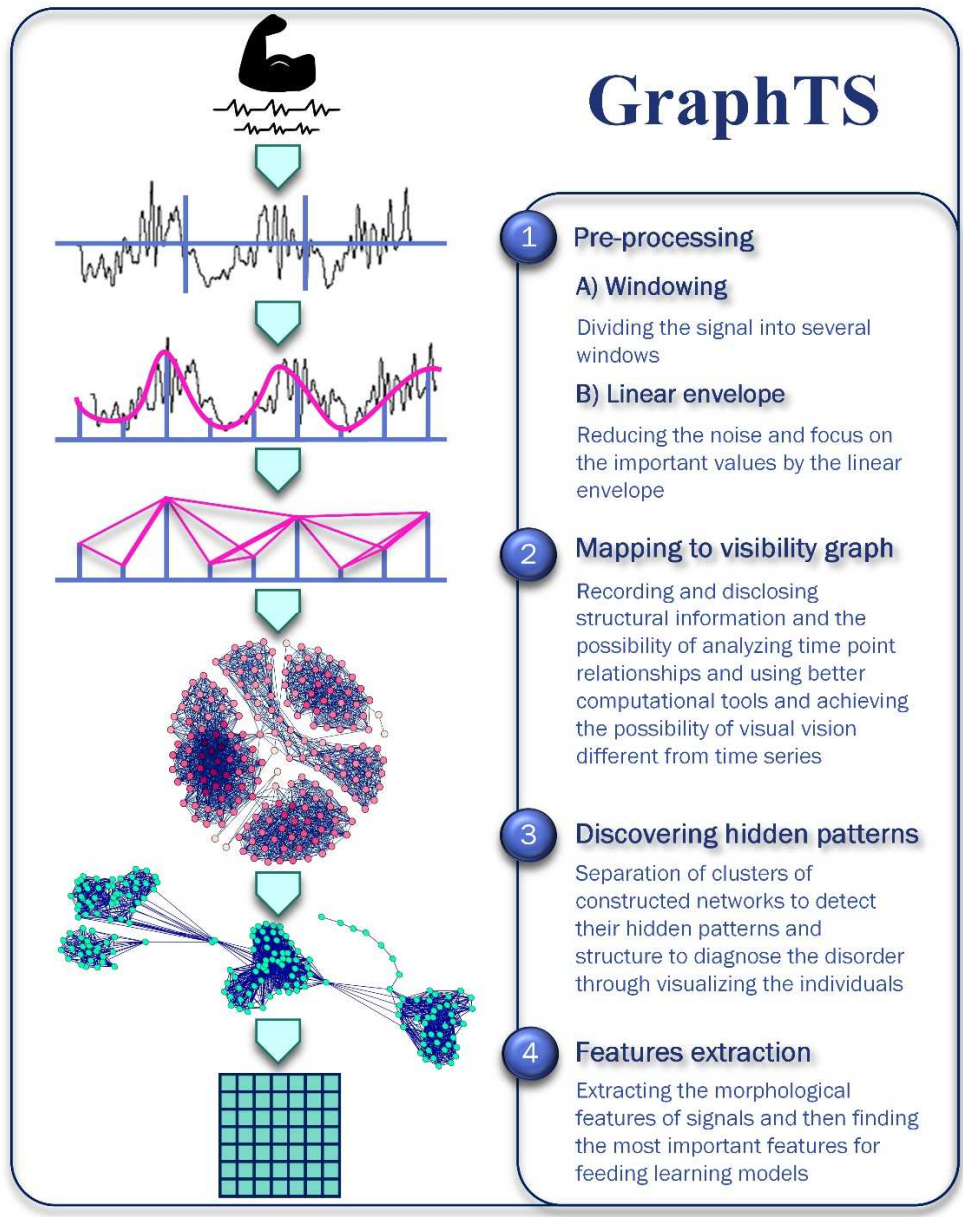

Figure 2. The overall process for the proposed GraphTS method

A portion of the electromyography signal of a patient with neuropathy is depicted in Fig. 3. A considerable number of signal points are located between several recorded noises and are separated from other noisy points by windowing. Windowing can be regarded as one of the primary stages of EMG signal preprocessing, which also plays an essential role in the initial steps of our method.

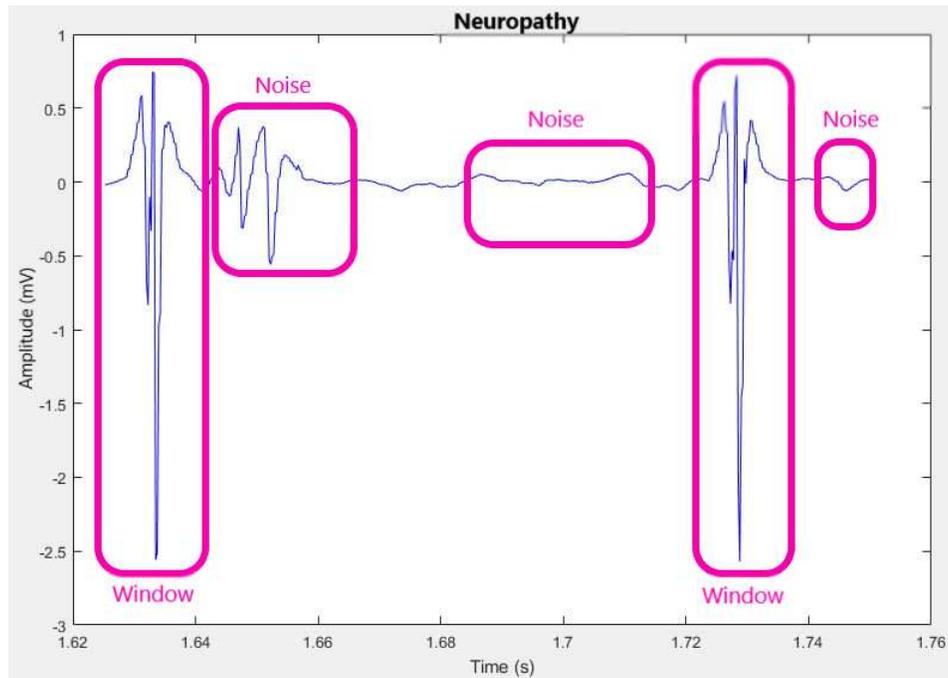

Figure 3. Part of a neuropathy electromyography sample, including noise and windows

### 4.1.2 Obtaining linear envelope of the signal

Initially, the absolute value of the signal is calculated. This step, called *full-wave rectification*, is performed to obtain an EMG signal envelope and is considered a fundamental operation in our method. It might be assumed that the signal envelope can be obtained by passing it through a low-pass filter even without rectification. However, this will not work well because the EMG signal fluctuates rapidly, and the rate of change of these fluctuations is around zero. So, this signal naturally has an average close to zero, and if it turns into a signal with a slight slope, the value obtained is almost zero, which is not applicable. If rectification is done first, the negative fluctuations turn into positive values. The result of passing the rectified signal through the low-pass filter in the range of 5 to 100 Hz looks like the main signal envelope.

Steps are taken to pass the signal through the low-pass filter. The first step is to average the signal in the desired window. This approach is called *rectification and averaging*. Getting a moving average of the signal is an example of a finite impulse response (FIR) filter. If this window is symmetrical and centered, it will not change the signal phase or timing. The filters that do not change the phase are called zero-phase shift filters. A discrete version of the low-pass filters, such as the Butterworth filter, is used to pass through the rectified signal low-pass filter. This is called the infinite impulse response (IIR) filter, often used in both forward and backward modes because the results have zero fuzzy shift. Combining the steps of rectification, averaging, and passing the low-pass filter is called the "finding linear envelope of the signal" approach [36].

In short, at this stage of preprocessing, the absolute value of the signals is first windowed and, then, the moving average is calculated to reduce the intrinsic fluctuations of the signal and focus on its signal's initial values. The linear envelope of the signal is then obtained by adjusting the parameters of the Butterworth filter and passing the signal values through it. The results of the action performed on a 0.025-second window of EMG of the healthy sample can be observed in Fig. 4. The green signal is the moving average of the absolute value of the signal, and the red signal is the linear envelope of the signal passed through the Butterworth filter.

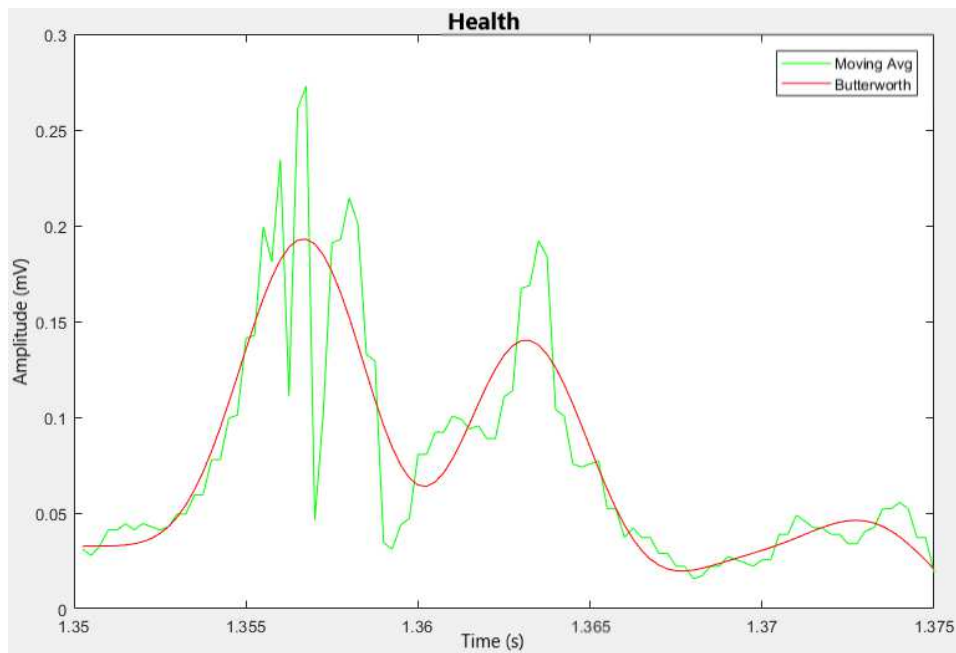

Figure 4. Step to find the linear envelope of signal

## 4.2 Time series mapping to the visibility graph

Complex network modeling plays a crucial role in discovering natural or social processes [37, 38]. Moreover, complex network analysis is an efficient approach for extracting hidden information in time series [14, 39]. Conventional time series analysis gives us information about the dynamics of a system [40]. However, network-based time series analysis provides a structural framework for modeling complex systems as well as their connections [41]. This section proposes to model the time series as a specific network architecture called a *visibility graph.* The reason for selecting this architecture will be discussed, followed by elaborating on the mapping of EMG signal to a visibility graph.

Complex networks have recently led to improved time series analytical approaches. The new methods are based on converting a time series to a network using specific algorithms of every type of mapping and extracting information from the resulting networks in the time domain [42]. On the other hand, interdisciplinary work in medicine usually acts as an assistant to an expert. Consequently, converting time series to complex networks gives the expert an entirely different view. A well-designed mapping will help the expert perform additional analyses on the resulting network and what the specialist understands from the time series itself. In this way, medical errors are reduced. The treatment path is better and faster due to a more accurate diagnosis of the disease and a comprehensive examination of its dimensions.

The development of a complex network enables the expert to work with computational tools with greater potential accuracy and newer approaches to time series analysis. Accordingly, the proposed method uses the standard visibility graph mapping method to convert EMG time series to complex networks. There are differences between the standard visibility graph and the *horizontal* visibility graph algorithms. The horizontal visibility graph is geometrically simpler and more analyzable than the standard visibility graph [30]. However, we do not use the horizontal type of visibility graph for EMG signals because the algorithm uses a horizontal connection, which fails to sufficiently distinguish the signal's characteristics. Some results of classification using the horizontal visibility graph are incorrect for EMG data [43].

## 4.3 Discovering hidden patterns

In a study on EMG data, the EMG signal has shown special patterns in its structure that include high peaks at specific times [30]. The horizontal visibility graph algorithm fails to detect this peak and specific signal patterns, so it may not be used to analyze this type of data. The standard visibility graph approach was chosen for mapping from time series to complex networks due to several reasons. It provides the ability to analyze the disease better, conforms to the nature of the signal, finds motifs in the network, provides interpretability and adaptation to the internal analysis. The visibility graph applies to fixed and non-stationary signals, tracks signal changes with efficient and straightforward calculations.

The next step is to implement the visibility graph approach. We used MATLAB for implementation. For each preprocessed window of the dataset, this algorithm generates a visibility graph of the signal line after receiving the linear envelope of the signal. After creating the desired visibility graph, its adjacency matrix will be given as output. Graph adjacency matrices developed for each signal are then used to represent the resulting complex networks and extract their features.

Samples of healthy, myopathy, and neuropathy signals are represented in a complex network using a visibility graph approach. This representation is made by Gephi software. For instance, Fig. 5 indicates the complex networks for the healthy, myopathy, and neuropathy groups. These networks correspond to windows with 200 time-points of EMG signals related to individuals. The color of the nodes of shown networks is adjusted and drawn by Gephi according to the degree of nodes from bold (node with the highest degree) to light (node with the lowest degree). The group of the signal can be found using the visualized network and multiple parts of each signal

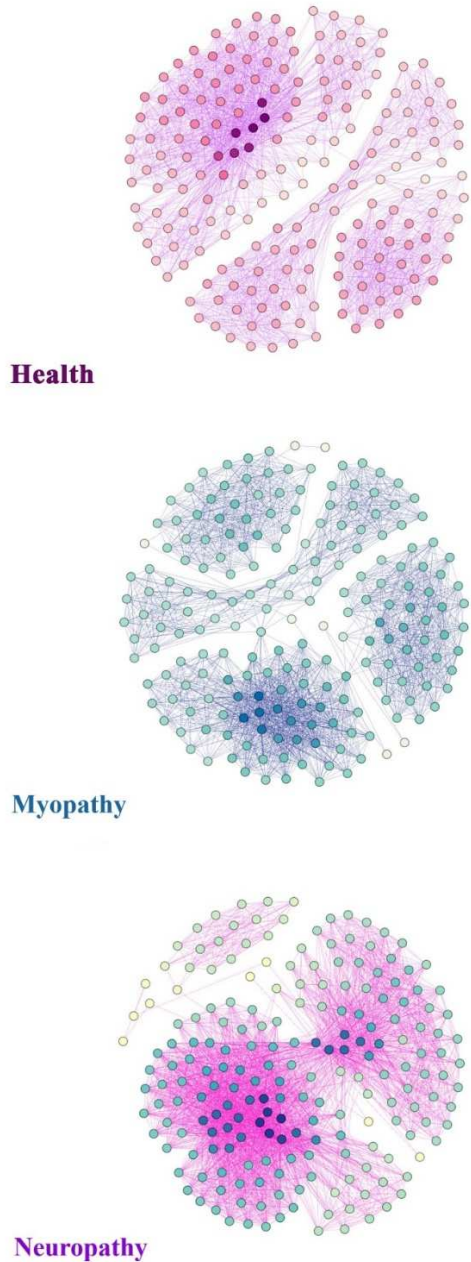

Figure 5. Visibility graph of a healthy sample (Upper), with myopathy (middle) and with neuropathy (down) with color change according to the degree of nodes

Subsequent steps to separate clusters and identify hidden structures in developed networks yielded significant results. It was performed after developing the networks with a visibility graph approach for all three groups with 200 time-points as nodes, 5292 edges for the healthy network, 4568 edges for the myopathy network, and 7178 edges for the neuropathy network, shown in Fig. 6. It can be seen that the EMG network of the healthy group (upper) with a blue graph, the EMG network of the myopathy group (middle), and the EMG network of the neuropathy group (lower) are shown with green and pink nodes, respectively. The upper (healthy) blue network is entirely interconnected, but the clusters have a separate branch in the middle green network (myopathy) and the lower pink network (neuropathy). Given the number of different clusters in each group and separate clusters in disease networks, all three networks can be wholly distinguished from each other.

Up to this point, the input data (windowing and signal envelope) were preprocessed. The complex network was then developed through the signal linear envelope calculated signal with the visibility graph approach. Besides, as shown in the figure, the networks of neuropathy and myopathy samples have a specific scheme and number of clusters. In general, the physician can determine whether the sample is healthy or diseased by observing the clusters. This will reduce medical errors and speeds up the analysis and diagnosis of the disease considerably. Following the correct and early diagnosis of the disease, the treatment process begins immediately. Lateral injuries are also prevented, and the patient will lose less time in the diagnosis process until treatment.

### 4.4 Feature extraction

In this step, the statistical and network features of the developed visibility graph are extracted using the adjacency matrix generated for each group. The features extracted from each network are stored after calculation as a feature vector consisting of the elements of the average degree [44], the average clustering coefficient [44], transitivity [45], density [46], network diameter [47], global efficiency [47], the average shortest-path [47]. After calculating the feature vector for each network, the vectors of all networks are put together to form a matrix of features extracted from the network. This matrix is given as input to the traditional machine learning models for classification.

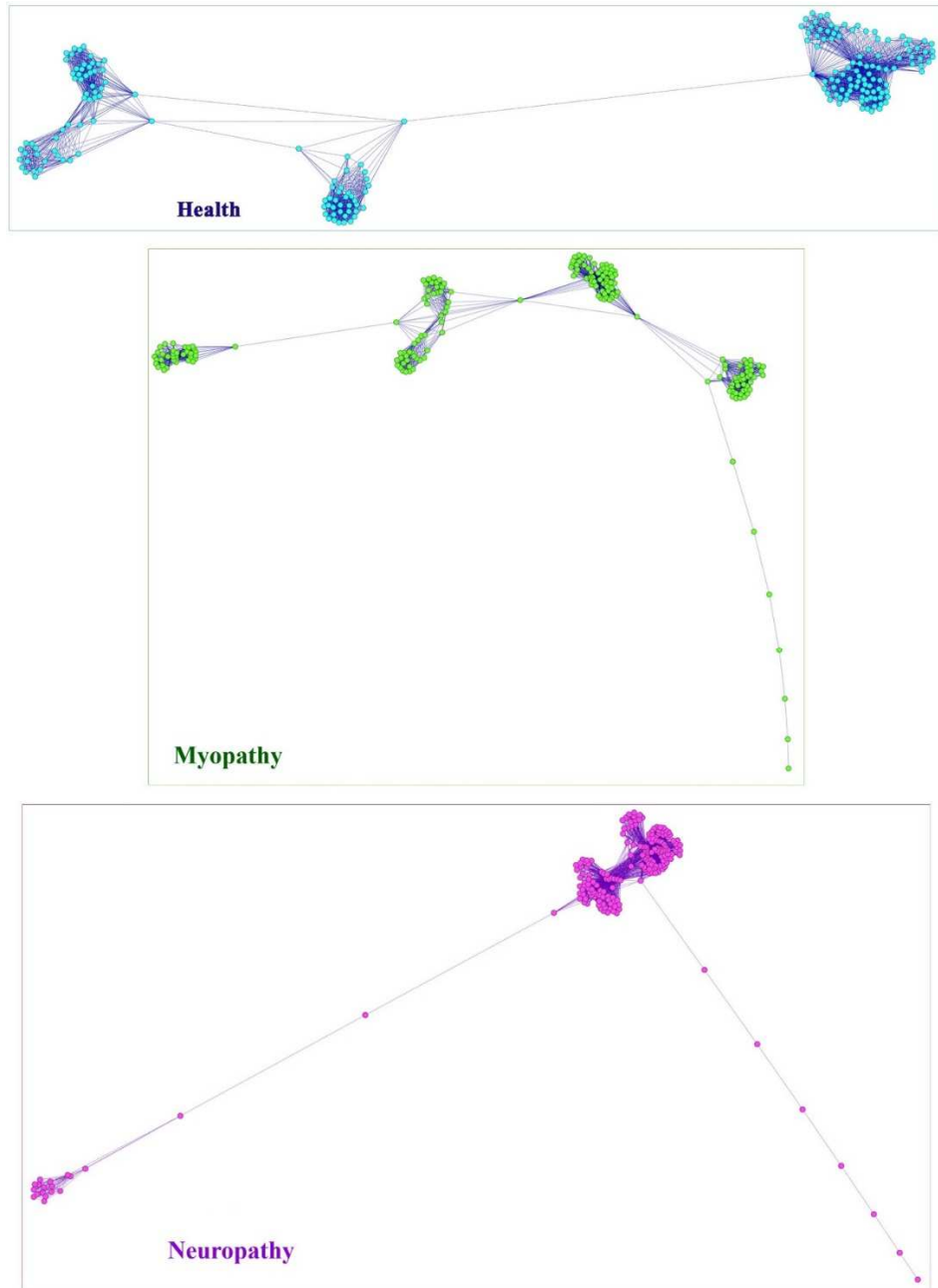

Figure 6. Visibility graph of a healthy sample (Upper), with myopathy (middle) and with neuropathy (down)

We applied the box plot to examine the distribution and skewness of extracted features, and then the best features have been selected for use in traditional machine learning models. For instance, Fig. 7 shows a plot for all the features extracted for the healthy group in blue (left), the neuropathy group in orange (middle), and the myopathy group in green (right). As shown in this plot, the distribution of features extracted from the network and the range of their data values are specified, a more detailed analysis of which is given in Section 5.

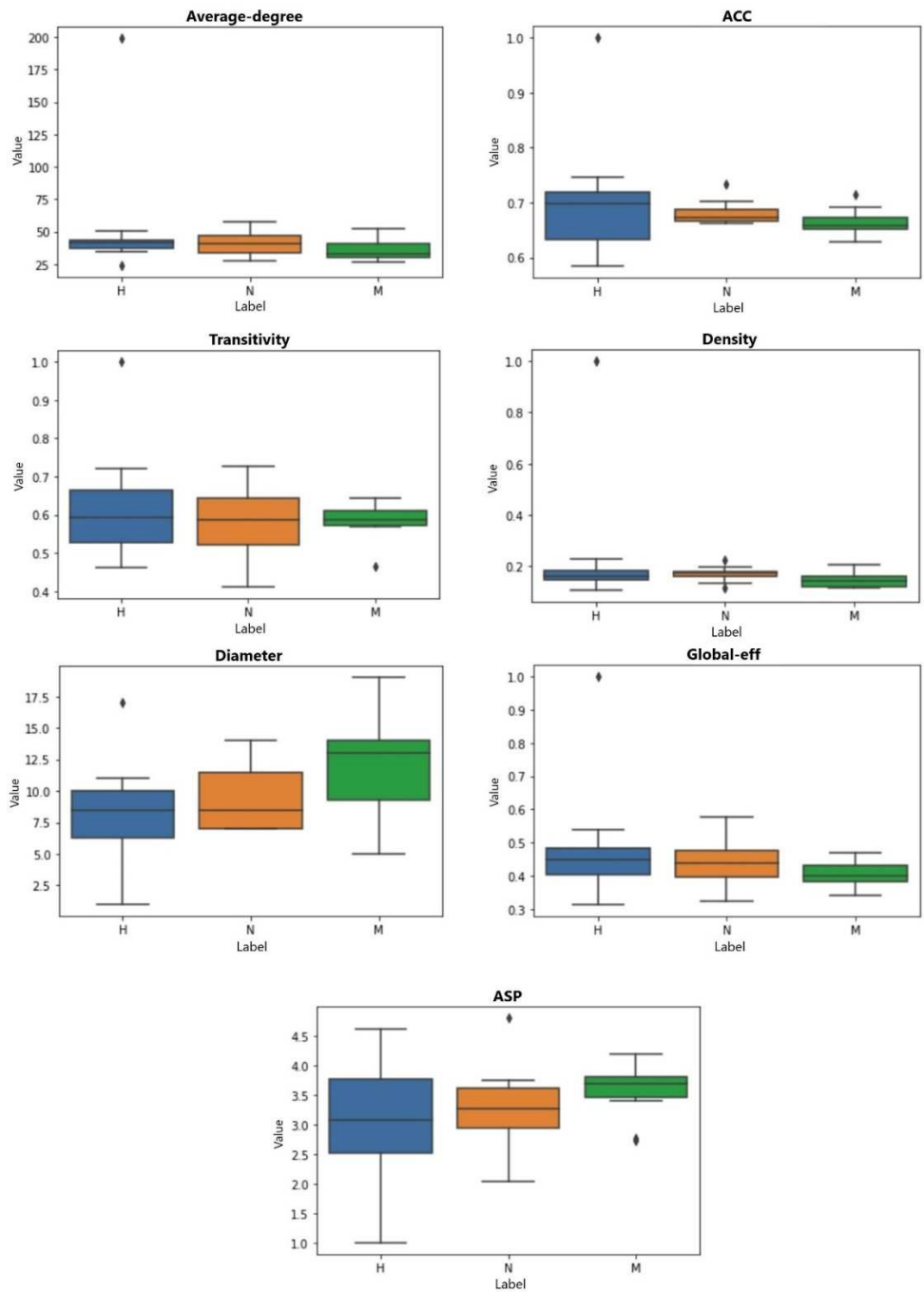

Figure 7. Box plot of network features distribution

## 4.5 Classification

Classification is one of the fundamental sub-disciplines of machine learning and data mining, which forms a model for predicting test data labels based on labeled training data. At this stage, our dataset is classified into three classes: healthy, myopathy, and neuropathy, using multiple machine learning algorithms to achieve the best possible results. These models include the k-nearest neighbor, random forest, logistic regression, support vector machine as traditional models and the artificial neural network, and deep neural network as novel models. As mentioned previously, the input data set (feature vector) is examined before applying it to traditional classification models using analysis of variance and the box plot to find the best features.

For the diagnostics and classification of diseases based on complex networks built from EMG data, three approaches exist based on proximity networks, visibility graphs, and transition networks. The authors in [28] presented VVASVM, another method in [29] called WVANN, and the NWVVA method in [30]. The results will be compared in Section 5.

## 5 Experimental evaluation

In this section, the EMG dataset and the hardware platform are described. Then the results of the experiments will be presented and discussed. The first set of experiments in Section 5.4 is focused on building a complex network from the EMG time series, followed by experiments on extracting the network features and training machine learning and neural networks for classification in Section 5.5.

### 5.1 Dataset

The first dataset [20] was collected by the Department of Neurology, Beth Medical Center, and Harvard Medical School using the Medelec Synergy N2 EMG signal recorder produced by the Oxford Medical Devices Research and Development Company. Three samples of EMG data were measured for healthy, neuropathy, and myopathy groups:

- A first example is a 44-year-old man with no history of neuromuscular disease;
- A second example is a 62-year-old man with chronic low back pain and neuropathy;
- A third example is a 57-year-old man with myopathy.

This dataset was recorded at a frequency of 50 kHz and then reduced to 4 kHz. Two analog filters (one 20 Hz high-pass filter and one 5 kHz low-pass filter) were used during the recording process.

The second dataset [25], the N2001 dataset, encompass three groups as follow:

- A group of healthy individuals
- A group of patients with myopathy
- A group of patients with ALS (a broad subset of patients with the neuropathy disorder)

The healthy group consisted of 10 people (four women and six men) in the age group of 37-21 years, six of whom were physically normal, except for one. None of the healthy groups had symptoms or a history of neuromuscular disorders. The group of patients with myopathy included seven or seven patients (two women and five men) in the age group of 19-63 years, all of whom had clinical signs of myopathy 15. The group of patients with neuropathy disorder (ALS) consisted of eight patients (four women and four men) in the age group of 35-67 years, five of whom, in addition to having neuropathy-compatible clinical signs, lost their lives due to the disorder within a few years of onset.

These two datasets are a combination of two different databases. Thus a more enriched dataset of EMG signals with three labels is formed: healthy group, myopathy group, and neuropathy group. Finally, we used the combined dataset with the mentioned specifications to train and test the implemented models.

## 5.2  Hardware specifications

The neural network is modeled, the deep neural network is applied on a system with GeForce GTX 1080 Ti graphics processing unit with Core-i7 processor and 32 GB of RAM, and the other models are simulated on a system with a Core-i5 processor and 4 GB of RAM to evaluate the classification models.

## 5.3  Evaluation measures

The classifier performance is evaluated by statistical measurements of accuracy, precision, recall, F-score, and specificity metrics. The accuracy is defined as the percentage of correctly classified divided by the number of all samples. F-score is computed as the harmonic mean of precision and recall, where precision is the number of actual positive samples divided by the number of all samples as positive. The recall is the number of actual positive samples divided by the total number of true positive samples. The values of precision, recall, and F-score can be calculated by Eq. (1) - Eq. (5), respectively.

$$Accuracy = \frac{TP + TN}{TP + FP + TN + FN} \quad (1)$$

$$Precision = \frac{TP}{TP + FP} \quad (2)$$

$$Recall = \frac{TP}{TP + FN} \quad (3)$$

$$F - score = 2 * \frac{Precision * Recall}{Precision + Recall} \quad (4)$$

$$Specificity = \frac{TN}{TN + FP} \quad (5)$$

### 5.4 Evaluation of complex networks

The GraphTS is a transparent approach that maps EMG time series to complex networks using a standard visibility graph algorithm step by step. It takes the fluctuating patterns into account and focuses on the initial values of the signal and map corresponding to healthy, myopathy, and neuropathy groups. The resulting networks inherit the main features of the raw signal. The network built by GraphTS can be used as a valuable tool for the physician along with the signal itself and play an influential role in the analysis and diagnosis of diseases.

As shown earlier in Section 4, Fig. 5, and Fig. 6, sample complex network clusters resulting from GraphTS are distinguished to identify the hidden structural pattern. Based on this structure, 200 time-points from the signals of the healthy, myopathy, and neuropathy groups, correspond to 200 nodes per network. The number of edges is 5292 for the healthy group, 4568 for the myopathy group, and 7178 for the neuropathy group. As shown in Fig. 6, the healthy sample network (top) is a fully interconnected network with three separate clusters. The myopathy sample network (middle) has four separate clusters, and the neuropathy sample network (lower) has two separate clusters. This difference between the number of resulting network clusters can be attributed to the unique fluctuating patterns of each signal set.

## 5.5 Feature assessment

The box plots in Fig. 7 are displayed separately for each feature with labels H for healthy, M for myopathy, and N for neuropathy. The data scatter rate, first and third quarters, medians, and outliers are also shown. Based on the plots and the values available, the network features with correct distribution and different medians for every data label will play a more notable role in classification.

When working with EMG data in our experiments (and in existing research [30, 43]), the peak points and different fluctuating patterns in specific time windows are crucial. So, the proposed approach emphasizes that the developed networks should focus on the signal's initial values. Therefore, it is expected that the features extracted from the resulting networks will also be affected. By examining the extracted features, it could be observed that the mean degree with medians of 44, 42, and 31.5, the average clustering coefficient with medians of 0.7, 0.68, and 0.66. The density has medians of 0.16, 0.174, and 0.11, and the shortest-path has medians of 1.3, 27.3, and 7.3 for healthy, neuropathy, and myopathy groups, respectively. With appropriate data scatter, can be considered as four essential and compelling features of the resulting networks.

The network features can be very effective in the classification of EMG datasets. The extracted features are also examined and analyzed for their variance before applying them to the support vector machine's two classifiers and the k-nearest neighbor. According to these studies, the classifiers receive and learn by those features that positively affect the classification results as input. The three features selected by analysis of variance from the seven features are the mean degree, the average clustering coefficient, and the network density.

## 5.6 Evaluation of classification models

The classification models are trained with 5-fold cross-validation. The dataset is divided into five subsets, four of which are used for training and one for validation in five rounds. The networks are developed from the EMG time series using the proposed GraphTS approach. It could be observed that there are distinguishable differences between healthy, myopathy, and neuropathy networks in the network structure. The network features are then fed into classification models, and the accuracy is calculated for each model.

In the following, the results of 20 repetitions for the k-nearest neighbor, random forest, logistic regression, and support vector machine in the classification into three classes of healthy, myopathy, and neuropathy groups are reported. The results of neural network and deep neural network classifiers are also reported for two-class and three-class classifications.

As earlier stated, all models are executed with seven network features as inputs, and the classification results are presented in Table 1. Among these seven features, four of them, such as *mean degree*, *average clustering coefficient*, *density*, and *average shortest-path* had a superior effect on the final performance of GraphTS. The classification results show that GraphTS favorably focuses on the initial values of each signal, peaks, and fluctuating patterns of each class. It effectively uses the visibility graph algorithm to construct a network that distinguishes healthy, myopathy, and neuropathy groups. The criteria for neural network and deep neural network models are presented in Table 1 based on the test data's accuracy. The accuracy of the training data for the ANN(II), DNN(II), ANN(III), DNN(III) was 96.14%, 99.30%, 94.50% and, 98.51%, respectively.

Table 1. Results of classification models with network features of our GraphTS approach

| Classification Model | Accuracy | Precision | Recall | F-score | Specificity |
| --- | --- | --- | --- | --- | --- |
| KNN | 0.5600 | 0.5600 | 0.5600 | 0.5500 | 0.6250 |
| SVM | 0.6148 | 0.6115 | 0.6100 | 0.6100 | 0.6660 |
| LR | 0.6931 | 0.8305 | 0.6050 | 0.6990 | 0.8148 |
| RF | 0.7185 | 0.8644 | 0.6296 | 0.7280 | 0.8518 |
| ANN(II) | 0.9470 | 0.8238 | 0.9684 | 0.8929 | 0.8939 |
| DNN(II) | 0.9918 | 0.9932 | 0.9865 | 0.9898 | 0.9953 |
| ANN(III) | 0.9180 | 0.8243 | 0.9682 | 0.8900 | 0.8910 |
| DNN(III) | 0.9738 | 0.9780 | 0.9707 | 0.9742 | 0.9855 |

The classification results are also shown in Fig. 8, sorted in ascending order by the accuracy. It could be observed that the k-nearest neighbor model achieves the minimum accuracy. The maximum belongs to the deep neural network model.

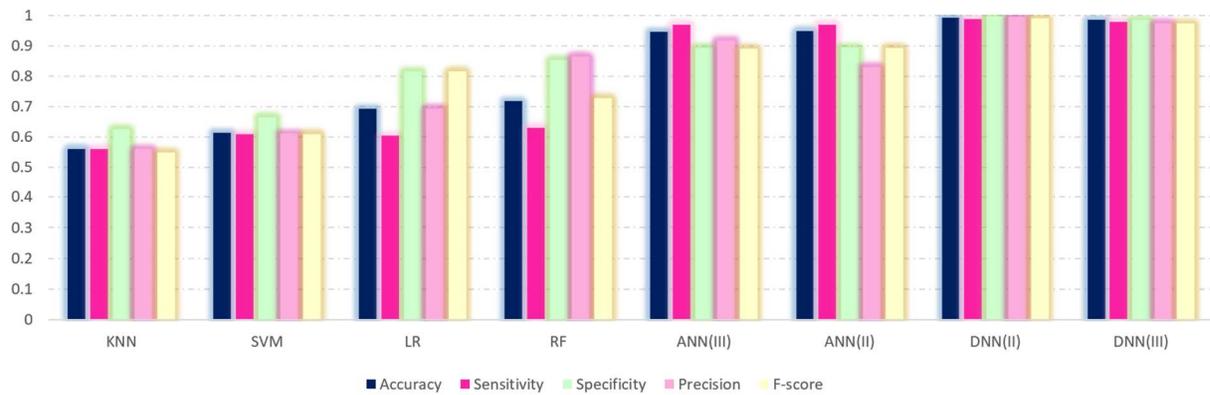

Figure 8. Results of classification models with the proposed GraphTS approach

When a graphic processing unit (GPU) is available, the highest classification accuracy could be obtained using the neural network and deep neural networks. In the absence of this hardware or limitation of time, the best model that will achieve the highest accuracy in the shortest time is a random forest. It provides a high degree of recognizability and accuracy compared to other classification algorithms.

The next part of the experiments examines the classification and diagnosis methods that were already mentioned in Section 4.5. The results are shown in Table 2. Higher values of accuracy are obtained by the VVASVM method.

Table 2. Comparing the classification and diagnosis results for three existing methods

|  | KNN | SVM |  | ANN |  |
| --- | --- | --- | --- | --- | --- |
|  | NWVVA | VVASVM | NWVVA | WVANN | NWVVA |
| **Accuracy** | 0.9462 | 0.9907 | 0.9831 | 0.9554 | 0.9750 |
| **Recall** | 0.9561 | 0.9810 | 0.9862 | 0.9624 | 0.9477 |
| **Specificity** | 0.9204 | 0.9953 | 0.9718 | 0.9415 | 0.9730 |

Similar criteria of the classification models for GraphTS approach can be seen in a window on a bar chart in Fig. 9. It could be observed that the desired accuracy is obtained for the neural network and deep neural network models. The key features of our approach have been focusing on the initial values, finding the linear envelope, dramatically reducing the noise in the initial signal, selecting the appropriate technique, and mapping time series to an enriched complex network using the visibility graph algorithm. Subsequent tasks could be summarized as feature extraction, selection of the desired feature through box plots, adjusting the weights of ANN layers, balancing weights and input datasets, and selecting the appropriate activation function and optimizer. This has led to improvements in both visualization and classification results.

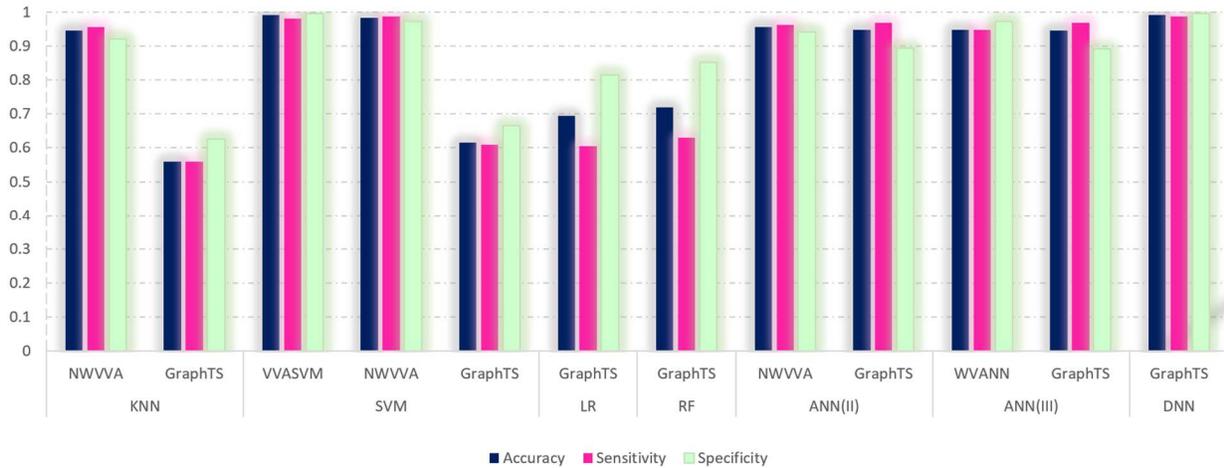

Figure 9. Comparing classification results

Moreover, the recall rate has increased significantly compared to other methods in addition to increasing accuracy. High recall is essential in the classification and diagnosis of diseases because the higher the recall, the less that model detects a patient group as healthy. This will significantly help to diagnose the disease quickly and to speed up the treatment process with the least waste of time.

# 6  Conclusion

In this paper, the GraphTS approach was proposed for constructing a complex network from EMG time series data, using standard visibility graph algorithm and EMG data of healthy, myopathy, and neuropathy groups. It was provided to diagnose and classify myopathy and neuropathy neuromuscular disorders both visually and through machine learning with network features. The EMG signals representing muscle responses to muscle stimulation and contraction were first transformed into a complex network using the visibility graph algorithm after windowing and obtaining the linear envelope of signals (to reduce fluctuations and focus on the initial values).

The novel visualized results show significant structural differences between the complex networks developed through the datasets of healthy, myopathy, and neuropathy. These networks can be interpreted in terms of having multi-part and structural differences based on the fluctuating patterns of the signals of healthy, myopathy, and neuropathy groups. The networks can be used to diagnose this type of disorder by physicians and experts in the field. This approach can be very effective in diagnosing neuromuscular disorders and accelerating the treatment of patients with these disorders.

In the next step, the network features were extracted by examining the adjacency matrix of the developed networks, which shows the relationship between the nodes of the resulting networks. Finally, the four extracted features were identified as the most effective ones in terms of data scatter and comparison of the medians of each group by box diagram and analysis of variance.

The extracted feature vector was then applied as input data to the support vector machine, k-nearest neighbor, logistic regression, random forest, neural network, and deep neural network classification models. The proposed GraphTS approach along with the deep neural network classification model achieved 99.30% accuracy on training data and 99.18% accuracy on test data, which is higher than all models in this study and previous studies. Besides, comparing the recall rate obtained, which is very important in the diagnosis and classification of diseases, with the proposed approach, indicated that this rate improved the diagnostic performance for healthy, myopathy, and neuropathy groups compared to other competitors' approaches.

Future works might include investigating other types of time series that are used in patient diagnosis, possibly leading to different topological and network features. Investigating other preprocessing and different network feature extraction methods also remains open and may lead to different classification and diagnosis results.